\documentclass[sn-basic]{sn-jnl}
\jyear{2022}%

\theoremstyle{thmstyleone}%
\theoremstyle{thmstyletwo}%

\theoremstyle{thmstylethree}%

\usepackage{amssymb}
\usepackage{epsfig}
\usepackage{multirow}
\usepackage{enumerate}
\usepackage{amsmath}
\usepackage{float}
\usepackage[section]{placeins}
\usepackage{url}
\usepackage{soul}

\begin{document}
\title[FRB luminosity distribution]{Luminosity distribution of fast radio bursts from CHIME/FRB Catalog 1 by the updated Macquart relation}

\author[1,2]{Xiang-Han Cui}
\author*[1,2,3]{Cheng-Min Zhang}\email{zhangcm@bao.ac.cn}
\author[1,2,4]{Di Li}
\author[1]{Jian-Wei Zhang}
\author[1,2,5]{Bo Peng}
\author[1,2,5]{Wei-Wei Zhu}
\author[6,7]{Richard Strom}
\author[8]{Shuang-Qiang Wang}
\author[8]{Na Wang}
\author[8]{Qing-Dong Wu}
\author[9]{De-Hua Wang}
\author[10]{Yi-Yan Yang}

\affil[1]{National Astronomical Observatories, Chinese Academy of Sciences, Beijing 100101, China}
\affil[2]{School of Astronomy and Space Science, University of Chinese Academy of Sciences, Beijing 100049, China}
\affil[3]{School of Physical Sciences, University of Chinese Academy of Sciences, Beijing 100049, China}
\affil[4]{NAOC-UKZN Computational Astrophysics Centre, University of KwaZulu-Natal, Durban 4000, South Africa}
\affil[5]{Key Laboratory of Radio Astronomy, Chinese Academy of Sciences, Beijing 100101, China}
\affil[6]{Netherlands Institute for Radio Astronomy (ASTRON), Postbus 2, 7990 AA Dwingeloo, the Netherlands}
\affil[7]{Astronomical Institute ‘Anton Pannekoek’, Faculty of Science, University of Amsterdam, 1090 GE Amsterdam, the Netherlands}
\affil[8]{Xinjiang Astronomical Observatory, Chinese Academy of Sciences, Urumqi, Xinjiang 830011, China}
\affil[9]{School of Physics and Electronic Science, Guizhou Normal University, Guiyang 550001, China}
\affil[10]{School of Physics and Electronic Science, Guizhou Education University, Guiyang 550018, China}


\abstract
{Fast radio bursts (FRBs) are extremely strong radio flares lasting several micro to milliseconds and come from unidentified objects at cosmological distances, most of which are only seen once.
Based on recently published data in the CHIME/FRB Catalog 1 in the frequency bands 400-800 MHz, we analyze 125 apparently singular FRBs with low dispersion measure (DM) and find that the distribution of their luminosity follows a lognormal form according to statistical tests.
In our luminosity measurement, the FRB distance is estimated by using the Macquart relation which was obtained for 8 localized FRBs, and we find it still applicable for 18 sources after adding the latest 10 new localized FRBs.
In addition, we test the validity of the luminosity distribution up to the Macquart relation and find that the lognormal form feature decreases as the uncertainty increases.
Moreover, we compare the luminosity of these apparent non-repeaters with that of the previously observed 10 repeating FRBs also at low DM, noting that they belong to different lognormal distributions with the mean luminosity of non-repeaters being two times greater than that of repeaters.
Therefore, from the two different lognormal distributions, different mechanisms for FRBs can be implied.
}


\maketitle

\section{Introduction}

Fast radio bursts (FRBs) are very strong radio sparks lasting micro to milliseconds in duration, which are mostly confirmed to be from objects at cosmic distances, which perhaps noticed in 1980 \citep{Linscott80}.
The FRB phenomenon was systematically studied by \cite{Lorimer07}, and then \cite{Thornton13} discovered several more sources in 2013.
Subsequently, this field developed rapidly \citep{Lorimer18, Zhang20, Petroff21}, including the first localized repeating FRB 121102 \citep{Spitler16, Chatterjee17}, the first FRB-like signal-FRB 200428 from the Galactic soft gamma repeater (SGR) 1935+2154 \citep{Bochenek20, CHIME20a, Lin20, Li21a} and FRB 20200120E in M81 \citep{Bhardwaj21}.
In addition, with further study, several research efforts showed that FRBs may be periodically active (FRB 180916: 16 day and FRB 121102: 157 day) \citep{CHIME20b, Rajwade20} and emit periodic signals (FRB 20191221A: 216.8 ms, FRB 20210206A: 2.8 ms, and FRB 20210213A: 10.7 ms) \citep{CHIME21b}.
Meanwhile, thanks to the completion of advanced radio instruments like the Canadian Hydrogen Intensity Mapping Experiment (CHIME) \citep{CHIME19a, CHIME19b, CHIME20a, CHIME20b, CHIME21a, CHIME21b}, Australian Square Kilometre Array Pathfinder (ASKAP) \citep{Shannon18, Kumar19},
and Five-hundred-meter Aperture Spherical radio Telescope (FAST) \citep{Li18, Zhu20, Luo20a, Lin20, Niu21, Li21b}, the data of FRBs, as well as their diverse properties, have dramatically increased.
For example, the 1652 and 849 bursts related to FRB 121102 were observed by FAST \citep{Li21b} and Arecibo \citep{Jahns22}, respectively, and CHIME released the CHIME/FRB Catalog 1\footnote{\url{https://www.chime-frb.ca/home}} \citep{CHIME21a} recently.

Before the CHIME/FRB Catalog 1 came out in June of 2021 \citep{CHIME21a}, the FRB Catalogue (FRBCAT)\footnote{\url{https://www.frbcat.org/}} was mostly used to study the statistical properties of FRBs \citep{Petroff16}.
Up to now, 129 FRB sources were published in the FRBCAT with various observation frequency bands from the different radio telescopes.
Although FRBCAT creates opportunities for us to pursue the matter of FRBs from a variety of frequency bands, the difficulties for statistical study are also confronted, such as the calibration of FRB data from the different telescopes and the uncertain spectral index between the diverse observational frequency bands \citep{Cui21a}.
Nowadays, after the CHIME/FRB Catalog 1 became available, the above dilemma is alleviated because data are only from the CHIME telescope at 400-800MHz \citep{CHIME21a}, and the amount of data has also increased significantly, 535 FRB data from 492 sources,
in which 462 apparent non-repeaters (hereafter, referred to as non-repeaters) are newly observed \citep{CHIME21a}.

The CHIME/FRB collaborations and other researchers have made many statistical analyses using the new database \citep{CHIME21a, Rafiei21, Chawla22, Pleunis21, Josephy21}.
Their conclusions further hint that the two groups of FRBs may originate from various paths, or they are likely to come from the same origin but different environments and conditions.
However, the FRB luminosity has not been discussed, perhaps because the source distance is hard to determine without localization.
In the former studies, luminosity function and distribution of FRBs have been discussed, and many of them are impressive and enlightening \citep{Kumar17, Li17, Luo18, Hashimoto20, Luo20b}.
Among the evidence and results, the luminosity distribution of FRBs is likely preferred to follow a power-law form or Schechter function as a whole \citep{Li17, Luo18, Lu19, Luo20b, Hashimoto22}, but this is still an open question because there exist a possibility of lognormal distribution \citep{Cui21a}.
Moreover, \citet{Li21b} recently found that the burst energy distribution of a repeating FRB 121102 is not a single power-law but a bimodal distribution, which may contain a lognormal component.
Besides these, whether the repeaters and non-repeaters share the same origin or physical properties is also a controversial issue\citep{Connor19, Caleb19}.
Some statistical analyses suggested that the different distributions for repeating and non-repeating FRBs may imply multiple origins or physical processes of FRBs \citep{Palaniswamy18, Petroff19, Fonseca20, Cui21b, CHIME21a}, while others believed that the difference is not so obvious or caused by the selection effects \citep{Connor20, Gardenier21}.
Therefore, according to these unsettle questions \citep{Kulkarni14, Petroff19, Cordes19}, by means of the new CHIME database,
the statistical explorations with various physical parameters between repeaters and non-repeaters can go ahead.

For the study of FRB luminosity, the distance estimation is a key step, where the Macquart relation between the dispersion measure and redshift (DM-z) \citep{Macquart20} is most useful, which is acquired by the 8 localized FRBs.
Here, as the first step, to test the validity of the Macquart   relation with the updated 18 localized FRBs, by adding the 10 new ones \citep{Heintz20}\footnote{\url{http://frbhosts.org/}}, we  showed that the Macquart relation is still applicable (details in Appendix A).
Therefore, we employed the Macquart DM-z relation and analyzed  the new CHIME data to estimate the FRB luminosity, 
processing the statistics that include the goodness of fit \citep{Cui21b}, Kolmogorov-Smirnov (K-S) test \citep{Smirnov48}, and Mann-Whitney-Wilcoxon (M-W-W) test \citep{Mann47}, Anderson–Darling (A-D) test \citep{Anderson52}, and Lilliefors test \citep{Lilliefors67}.

The structure of our paper is organized as follows.
In Section 2, we describe the data selection and luminosity estimation.
In Section 3, the FRB luminosity distribution is fitted by the different types of functions, and the results of statistical tests are given.
In Section 4, we discuss the selection effects, the error of Macquart  DM-z relation, the applicability of lognormal distribution, and the implications of statistical results.
Finally, a brief summary is exhibited at the end of the paper.

\section{Data selection and estimation}\label{2}

In this section, we elaborate on the data selection of CHIME/FRB Catalog 1 and the estimation of FRB distance, based on which the FRB luminosity is obtained.

\subsection{Data selection}

Our data are taken  from the first release of CHIME/FRB Catalog 1 \citep{CHIME21a} and its website\footnote{\url{https://www.chime-frb.ca/home}}.
This catalog contains 474 non-repeaters and 18 repeaters, of which 462 non-repeaters are published for the first time.
However, due to instrument responses, processing methods, and other potential effects, these data need selection for non-repeaters for the later analyses, which were mentioned in \citet{CHIME21a} and illustrated as follows.
\begin{itemize}
\item Events with the tab of "excluded\_flag = 1" should be removed, considering that these data may be influenced by non-nominal telescope operation.
\item Signal-to-noise ratio (SNR) should be larger than 12,  because events with SNR $<$ 12 are sensitive to human bias and are more likely to be misclassified as noise or radio frequency interference (RFI).
\item DM needs to meet the conditions that $\rm DM>1.5DM_{MW}$ and $\rm DM>100\,pc\,cm^{-3}$, to ensure these events are truly extragalactic sources, and where $\rm DM_{MW}$ we used is $\rm DM_{YMW16}$ model \citep{Yao17}.
\item Scattering timescale for events needs less than 10 ms, due to the selection effects bias against wide bursts.
\item Events detected in far side-lobes (indicated by the tab ``Flux note" of Catalog 1) were rejected, as reliable flux or position measurements are not yet available for these events.
\item Some sources have sub-bursts, so we only consider the first burst to avoid repetitive counting.
\end{itemize}

After these criteria, 255 non-repeaters were reserved.
However, the direct distance data have not been given for sources.
If we assume that the Macquart relation \citep{Macquart20} of DM and redshift (DM-z) correlation is correct, we can derive an empirical formula to roughly estimate the FRBs distance.
In addition, we also re-examine the Macquart DM-z relation by the latest updated 18 localized FRBs and find that it still follows the previous results, the detail of which is shown in Appendix A.
Then, another problem arises that the Macquart relation can only estimate the distance at low redshift or DM \citep{James22}.
Meanwhile, in Catalog 1, the the inferred power-law index for the cumulative fluence distribution of low DM (ranges in $\rm 100-500\,pc\,cm^{-3}$) and high DM ($\rm DM>500\,pc\,cm^{-3}$) is different \citep{CHIME21a}, where DM is the data that removes the Milky Way galaxy contribution.
Besides that, for high DM data, bright bursts are more likely to be recorded by receivers, which means that the distant faint burst is incomplete for the full sample.
Therefore, finally 125 non-repeaters with low DM are selected in our sample, as well as 10 repeaters at low DM.

\subsection{FRB Luminosity estimation}

We employed the Macquart relation to estimate the distance of FRB, so the first step is to clarify the DM we used.
FRB's DM is consist of 4 parts \citep{Cordes19}, originating from the Milky Way galaxy and halo ($\rm DM_{MW}$), intergalactic medium in cosmic distance ($\rm DM_{IGM}$), host galaxy ($\rm DM_{host}$) and surrounding medium ($\rm DM_{sur}$), respectively.
Because only 19 FRBs were localized, we know little information about other  host galaxies and their $\rm DM_{excess}$
($\rm DM_{IGM}$, $\rm DM_{host}$, and $\rm DM_{sur}$), where  $\rm DM_{excess}$ is used to estimate the upper limit of redshift and distance  and $\rm DM_{MW}$ is subtracted according to YMW model \citep{Yao17} but not NE2001 \citep{Cordes02}.
Here, we do not estimate the DM contribution of the Milky Way galaxy halo, because they are not well constrained by current observations so far \citep{Rafiei21}, which have a rough range of $\sim 10-100\,pc\,cm^{-3}$.
Since our DMs are all larger than $100\,pc\,cm^{-3}$ after the above selections, we can ensure that these sources are from extragalactic distances.
Thus, based on our method, the luminosity we calculate is the upper limit of the intrinsic value.

By using the basic knowledge of distance measures in cosmology \citep{Hogg99}, the proper distance ($d_p$) can be roughly estimated as $d_{p} = zc/H_0$,
where c is speed of light and $H_{0}$ is Hubble constant cited from \cite{Planck16} and \cite{Macquart20} ($H_{0}=\rm 67.74 \, km\,s^{-1}\,Mpc^{-1}$).
Based on the simplest assumption of a flat universe ($\Omega_k = 0$) and the definition of comoving distance ($d_c$),
the luminosity distance ($d_L$) is  written as $d_L = (1+z)d_{c} = (1+z)d_{p}/(1+z) = d_{p}$.
Meanwhile, the data of Catalog 1 are only from CHIME at the frequency band of 400-800 MHz,
so we do not need to consider the impact of the different telescope calibrations on the data.
Thus, combined with the FRB flux ($S$), the upper limit of luminosity of non-repeaters can be calculated as  $L \sim Sd_L^2/(1+z)$.
For the 10 repeaters, we also estimated their distances by the above method, and the first flux values of multiple observations are taken to represent this source. 

\section{Analysis and results}\label{3}


\begin{figure}
\centering
\includegraphics[width=6cm]{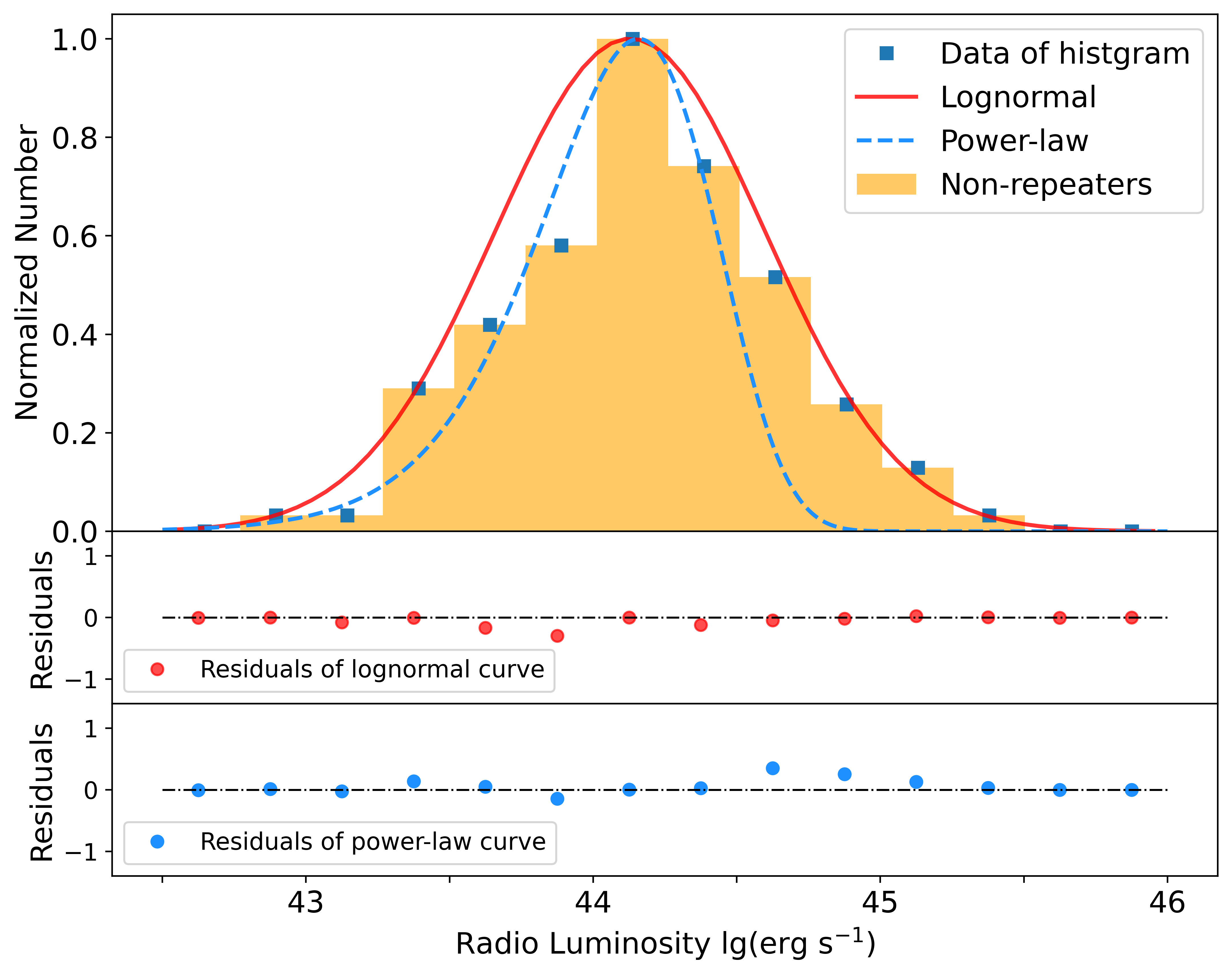}
\caption{
Upper panel: normalized histogram and different fitted curves of non-repeaters at low dispersion measure (DM) from CHIME/FRB Catalog 1.
The solid line is the curve of lognormal form and the dashed line is the curve of power-law type.
Middle and bottom panels are the residuals for the lognormal and power-law curve, shown as solid points, which are calculated by the normalized histogram data and curves.
The dash-dotted lines refer to the cases of residual = 0.
}
\label{fig1}
\end{figure}

In our former work about the luminosity distribution of FRBs, we elucidated that the lognormal distribution is better than the power-law type based on the FRB data by CHIME and other telescopes \citep{Cui21a}. 
However, our conclusion was weakened due to the small amount of data and the difficulty of calibrating between the different telescopes.
Nowadays, thanks to the CHIME Catalog 1,  we can employ sufficient and more uniform data to do the further statistical tests,  where the 125 non-repeaters mentioned above are applied and the histogram of their luminosity is plotted in Figure 1.
Meanwhile, 5 statistical methods are used to judge our results, including goodness of fit, K-S test, M-W-W test, A-D test, and Lilliefors test.
The analyses of full sample (255 non-repeaters) are shown in Appendix B.

To start our statistics, the test criteria need to be briefly introduced.
The test results are represented by the parameters of coefficient of determination ($R^2$) and p-values that are given under a 5\% significance level, and the procedures are as follows.
\begin{itemize}
\item For the goodness of fit, the closer $R^2$ is to 1, the better the fit. 
On the contrary, the fit is worse.
\item For the K-S test, M-W-W test, and Lilliefors test, the lower the p-value, the greater the difference from the null hypothesis. 
If the p-value is less than 0.05, it indicates that this test rejects the null hypothesis.
\item For the A-D test, when the value of the statistic is larger than the critical value, the null hypothesis can be rejected.
\end{itemize}
Firstly, we directly test the lognormal distribution by applying the K-S test, Lilliefors test, and A-D test, and the null hypothesis is that the sample is consistent with a lognormal form.
The results are shown in Table 1.
Meanwhile, all p-values and statistics are larger than 0.05 and critical values, respectively, which indicate that the lognormal form is consistent with both repeaters and non-repeaters under these three tests.
\begin{table}
\centering
\caption{Statistical test results of lognormal distribution for repeaters and non-repeaters at low DM}
{
\begin{tabular}{@{}lcccc@{}}
\hline
\noalign{\smallskip}
\bf FRBs &\bf Number &\bf K-S test &\bf Lilliefors test &\bf A-D test$^a$ \\
\hline
\noalign{\smallskip}
Non-repeaters & 125 & 0.872 & 0.605 & (0.333, 0.764)\\
Repeaters & 10 & 0.640 & 0.154 & (0.519, 0.684)\\
\hline
\end{tabular}
}
\label{tab1}
\begin{flushleft}
Notes. $^a$ The results of A-D test are shown in the form of statistic and critical values (statistic, critical values) for a 5\% significance level.
\end{flushleft}
\end{table}

Then, we use the lognormal and power-law types to fit the luminosity distribution (in Figure 1) 
and utilize the statistical methods to test the fitting closeness, including the goodness of fit, K-S test, and M-W-W test.
To avoid deviations caused by different coefficients, we normalized the data and fitted curves and performed statistical analysis on them.
From Figure 1, we can see that lognormal curve is closer to the data histogram comparing with power-law curve, and the fitting residuals of lognormal are also smaller than that of power-law, where power-law form is mentioned as Schechter function \citep{Luo18}.
From Table 2, the goodness of fit of lognormal ($R_{log}^2$ = 0.99989) is higher than that of power-law ($R_{power}^2$ = 0.820).
Moreover, further tests  of the K-S and M-W-W  on the fitting property also show that the lognormal conforms better than the power-law.
Although the goodness of fit of power-low is high, it is still smaller than that of lognormal, and p-values of power-law are all smaller than 0.05 in Table 2.
Therefore, from the distribution pattern and the above statistical test results, a preliminary conclusion is that the lognormal form is preferred for luminosity distribution.
\begin{table}
\centering
\caption{Statistical test results of different luminosity distribution for non-repeaters at low DM}
{
\begin{tabular}{@{}lccc@{}}
\hline
\noalign{\smallskip}
\bf Different types&\bf Goodness of fit &\bf K-S test  &\bf M-W-W test \\
\hline
\noalign{\smallskip}
Lognormal & 0.99989 & 0.630 & 0.795 \\
Power-law & 0.820 & $3.70\times 10^{-3}$ & $7.78\times 10^{-3}$\\
\hline
\end{tabular}
}
\label{tab2}
\begin{flushleft}
\end{flushleft}
\end{table}

However, we noticed that the above analysis is based on the strict Macquart relation between DM and z.
Considering that this relation is estimated from broad scattered data points, DM and z relation is not an exactly proportional one-to-one correspondence \citep{Luo20b, James22}.
That is, a DM may correspond to a distribution of z rather than a single value.
So, we introduce various errors to the DM-z transition for 125 non-repeaters to simulate this uncertainty.
The value of error injection is z, which is from 10\% to 100\% with a 10\% interval in the uniform and normal error pattern, respectively.
Specifically, error pattern means that the transition error of DM-z may belong to a particular distribution (uniform or normal), and its statistical characteristics (mean, upper and lower limits for uniform distribution, and mean and variance for normal distribution) are affected by error percentage (10\%-100\%).
Therefore, for each error percentage and error pattern, we perform the Monte Carlo sampling 300 times based on two assumptions:
the first is that the probability and shape of error in each data set is the same in one sampling;
secondly, the sampling range for each data set depends on its own characteristics.
For example, for the data with z = 0.5 and a 20\% normal error shape, we take 300 times sampling within normal distribution with a mean of 0.5 and a variance of 0.1 ($0.5\times20\%=0.1$).
It should be noted that for the situation of large errors like 100\%, values of 0 may occur during the sampling process, for which we only use values greater than 0 for analysis.

Then, we find the z with maximum (z-max) deviations that are compared with their averages in each group.
After that, the values of luminosity are calculated by using the above z-maxs drawn in Figure 2 (uniform) and Figure 3 (normal), and we also plot the original lognormal fitting curve (the red line in Figure 1) in each sub-figure.
Meanwhile, K-S, M-W-W, and Lilliefors tests are applied for each group's luminosity to check whether they are still consistent with the lognormal form.
Nevertheless, in fact, this only performs statistical tests on one set of data, which may cause fluctuations.
Thus, to eliminate the randomness of sampling, we do a 100 times loop for the above sampling, and take the harmonic mean values for each $p_{ks}$, $p_{mww}$, and $p_{lilliefors}$ individually to represent the test results under the corresponding error percentages and patterns, which is listed in Table 3.
The further discussions are shown in Section 4.2.

\begin{table}
\centering
\caption{Statistical test results (p-values) of luminosity distribution with various error shapes and error percentages}
\resizebox{\textwidth}{15mm}
{
\begin{tabular}{@{}lccccccccccc@{}}
\hline
\hline
\noalign{\smallskip}
\multirow{2}{*}{\bf Statistical tests } &\multirow{2}{*}{\bf Error shapes } & \multicolumn{10}{c}{\bf Error percentages}  \\
                                &            & 10\%   & 20\%   & 30\%   & 40\%   & 50\%   & 60\%   & 70\%   & 80\%    & 90\%     & 100\%   \\
\hline
\noalign{\smallskip}
\multirow{2}{*}{\bf K-S test}   & Uniform    & 0.748  & 0.746  & 0.780  & 0.800  & 0.796  & 0.762  & 0.697   & 0.499   & 0.241   & 0.0128 \\
                                & Normal     & 0.794  & 0.811  & 0.637  & 0.204  & 0.168  & 0.225  & 0.0422   & 0.0416   & 0.0334   & 0.0782  \\
\multirow{2}{*}{\bf M-W-W test} & Uniform    & 0.889  & 0.878  & 0.870  & 0.868  & 0.855  & 0.848  & 0.839   & 0.784   & 0.651   & 0.303 \\
                                & Normal     & 0.884  & 0.873  & 0.804  & 0.602  & 0.528  & 0.539  & 0.434   & 0.402   & 0.411   & 0.447   \\
\multirow{2}{*}{\bf Lilliefors test} & Uniform    & 0.363  & 0.323  & 0.391  & 0.374  & 0.421  & 0.340  & 0.235   & 0.0509   & 0.00949  & 0.00144 \\
                                & Normal     & 0.416  & 0.448  & 0.108  & 0.00975  & 0.00683  & 0.00967  & 0.00328   & 0.00247 & 0.00255 & 0.00334  \\                                
\hline
\hline
\end{tabular}
}
\label{tab3}
\begin{flushleft}
\end{flushleft}
\end{table}

\begin{figure*}
\centering
\includegraphics[width=11cm]{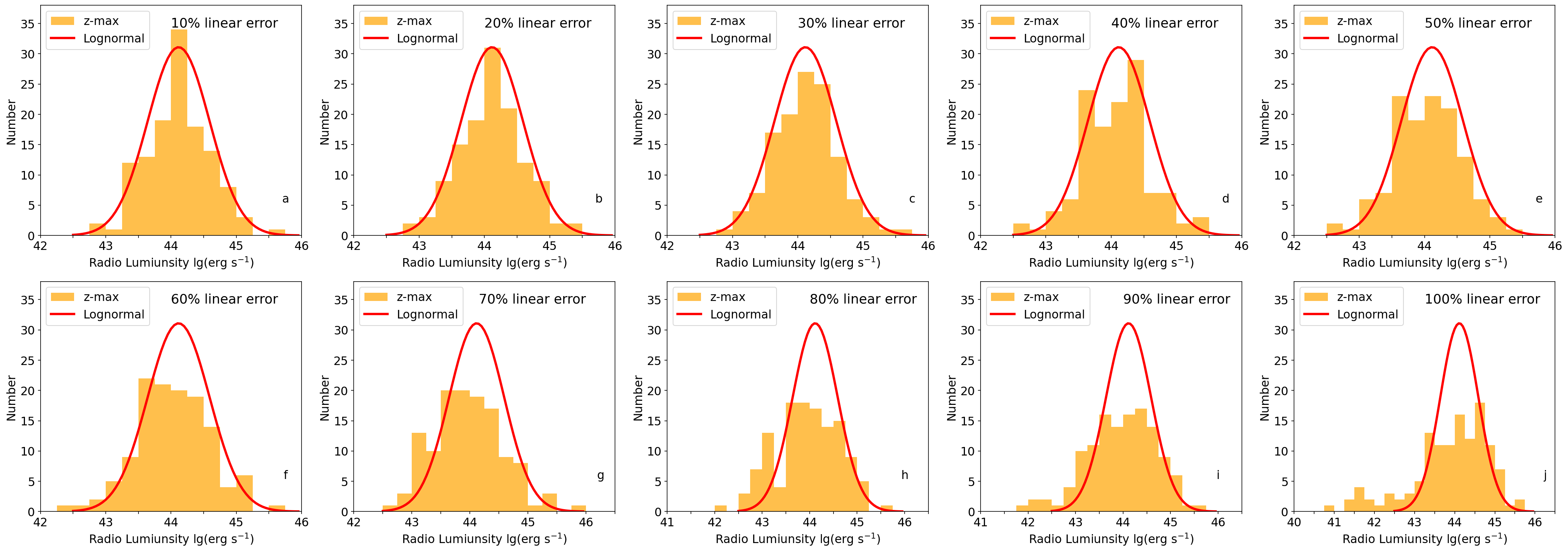}
\caption{Luminosity histogram of non-repeaters at low dispersion measure (DM) in uniform error shape with changing error percentages from 10\% to 100\%, from sub-figure a to sub-figure j.
In each sub-figure, the cross-hatched histogram is the maximum deviation and the empty one means the minimum deviation of 300 times sampling
in the corresponding error percentage.
The line represents the best fitting curve without considering the error of the Macquart relation, which is an unnormalized fitting curve in Figure 1.}
\label{fig2}
\end{figure*}

\begin{figure*}
\centering
\includegraphics[width=11cm]{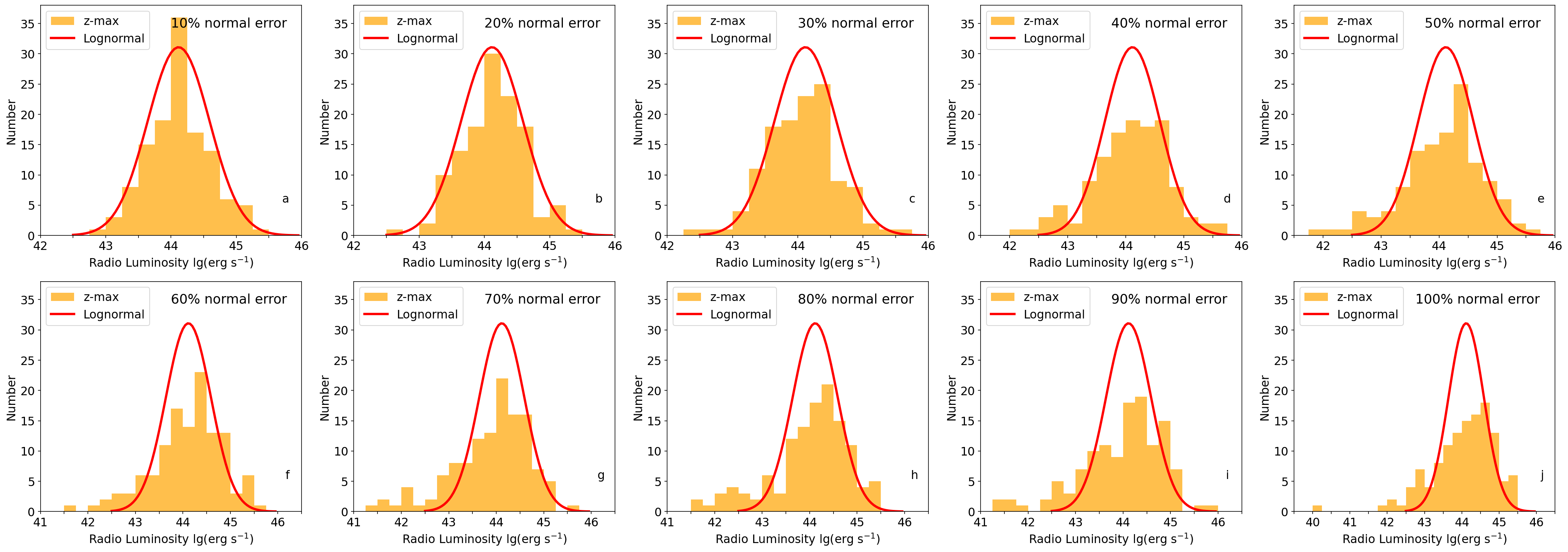}
\caption{Luminosity histogram of non-repeaters at low dispersion measure (DM) in normal error shape with changing error percentages from 10\% to 100\%,  from sub-figure a to sub-figure j.
Other annotations are consistent with those in Figure 2.}
\label{fig3}
\end{figure*}


\begin{figure}
    \centering
    \includegraphics[width=6cm]{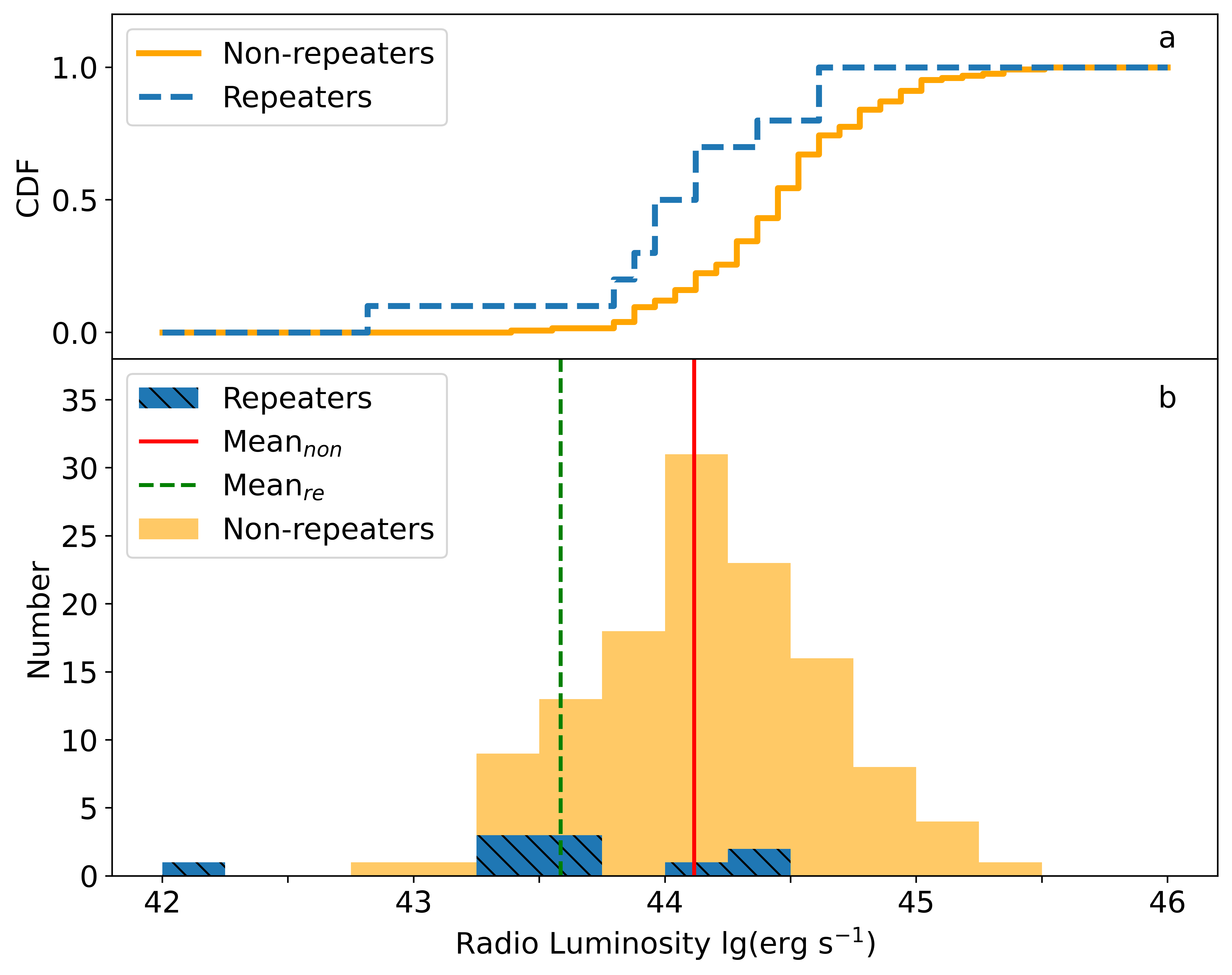}
    \caption{The distinction between the repeaters and non-repeaters at low dispersion measure (DM) in the aspect of luminosity.
    The top panel (sub-figure a) is the CDF of repeaters and non-repeaters for luminosity.
    The dashed (solid) line is for repeaters (non-repeaters).
    The bottom panel (sub-figure b) is the histogram of luminosity for two samples.
    The cross-hatched (empty) histogram means the repeaters (non-repeaters).
    The dashed (solid) line represents the mean value of repeaters (non-repeaters).}
    \label{fig4}
\end{figure}

Finally, we analyze the data of repeaters and non-repeaters.
Because the luminosity distribution of repeaters had been discussed in our previous work and it also showed the lognormal \citep{Cui21a}, we will not make the redundant analysis of the luminosity distribution for the repeaters here.
But, it is necessary to compare the distributions of repeaters and non-repeaters again, by considering that there were only 12 non-repeaters in the comparison of the previous work.
Therefore, we compare the two samples of repeaters (10) and non-repeaters (125) at low DM in the aspect of luminosity, the distributions of which are plotted in Figure 4, and we employ the K-S and M-W-W tests on these two samples.
The results are obtained as $p_{ks}$ = 0.0120 and $p_{mww}$ = 0.00943, indicating that the two samples may follow the different statistical distributions.
These results infer that, although their distribution types are the same, the specific statistical characteristics are different.

\section{Discussions and conclusions}\label{4}

Four aspects will be discussed in this section: selection effects, the applicability of the lognormal for FRB luminosity, the magnetar origin for FRBs, and the difference between the repeaters and non-repeaters.
To begin with, we need to clarify the possible selection effect in our sample, discussing whether the lognormal distribution will be varied as the error percentage of DM-z relation changes.
Then, we discuss what kinds of magnetars can produce the FRB luminosity as obtained, and the models for the repeaters and non-repeaters.

\subsection{Selection effects}
The detection position from far side-lobes and off-axis beams will effectively affect the performance of the telescope.
Although most of these data have been removed after selection, we cannot guarantee that all flux data are under the same instrument response due to the current positional uncertainties of the CHIME/FRB Catalog 1 events.
But we believe that this may be a minority case and has less impact on the conclusions.
Besides that, we should note that the data of burst flux in CHIME/FRB Calalog 1 are biased, such as in high-DM and wide bursts with long scattering time, so \citet{CHIME21a} injected simulate bursts to quantify this effect.
However, we do not use this method but apply the data selection of low-DM and short scattering events to study the FRBs population.
Therefore, it should be remarked that our analyses are only applicable to the current selection method. 

Due to repeaters having been observed several times, they may obtain better measured DM data and estimated fluence.
These effects probably cause differences in selection functions compared with non-repeaters samples, we cannot use this statistical result to determine whether a particular source will be repeated in the future.
Thus, the difference between them may stem from selection effects like beaming and propagation, or the intrinsic characteristics of FRBs are various \citep{Pleunis21}, and our following discussion mainly focuses on the possibility of intrinsic features.

\subsection{Applicability of the lognormal distribution}
Because our study on the FRB luminosity depends on the distance estimation by the Macquart DM-z relation, which needs further discussions as below.
To verify the Macquart relation, we firstly use the new data of  18 localized FRBs to confirm its validity as previously declaimed (as shown in Figure 5 in Appendix A).
We obtain that the slope (k)  by using a linear function for fitting ($k_{18}\sim 1028$ ) is almost parallel with the value of Macquart relation ($k_m\sim 973 $), and the difference between the two slops is no more than 6\%.
Under the constraints of 10 new data of localized FRBs, the Macquart relation does not change significantly, and the further analysis is listed in Appendix.
Therefore it is reasonable that we employ this DM-z relation to estimate the distance to calculate the FRB luminosity. 
Then, the approximated isotropic luminosity data of FRBs based on the CHIME/FRB Catalog 1 are acquired under the Macquart relation, which follows the lognormal distribution and rejects the power-law type.

However, considering that the Macquart relation is not a strict one-to-one conversion from DM to z (distance), the lognormal distribution of FRB luminosity perhaps to be deformed with the increase of error for DM-z correlation.
Here, with the different error shapes (uniform and normal type) and error percentages (from 10\% to 100\%), the variations of the lognormal characteristics are represented by statistical test results ($p_{ks}$, $p_{mww}$, and $p_{lilliefors}$), which means that the lognormal distribution may be constrained with a certain scope of application. 
According to Table 3, for the different error percentages, the lognormal characteristics are gradually decreasing under K-S, M-W-W, and Lilliefors test results.
Although the majority of p-values are greater than 0.05, this clear downward trend indicates a weakening of lognormal feature.

Specifically, for the uniform error shape, when the percentages reach 100\% in the K-S test and 90\% in the Lilliefors test, the corresponding p-values are lower than 0.05.
For the normal error shape, the p-values smaller than 0.05 have appeared while the percentages go to 70\% and 40\% in K-S and Lilliefors test, respectively.
Therefore, the lognormal feature is threatened in the above situations.
The reason for the different p-values under multiple tests lies in the different emphasis of each method.
For example, the K-S is tests the maximum distance between an empirical distribution function and a cumulative distribution function, while the M-W-W analyzes by the median.
Lilliefors test is an improved K-S test, which is appropriate when the test distribution must be specified by parameters estimated from the data \citep{Lilliefors67}.
This also explains why $p_{lilliefors}$ always precedes $p_{ks}$ less than 0.05.
If we give the criterion that they pass all three tests in both error shapes, they will meet the lognormal distribution.
Thus, it can be inferred that when the estimation error of the Macquart relation between DM and z is less than 40\%, the lognormal luminosity distribution is credible.
On the contrary, the lognormal feature is not obvious.

It should be clarified that the errors we are discussing here do not come from a particular uncertainty, but representing all possible aspects in DM-z transition.
This is intended to simulate and test whether the form of luminosity distribution changes when uncertainty is introduced, rather than to describe and discuss the type and cause of specific error.
Indeed, the selective effect and the observational bias may significantly affect DM-z conversion,
so \cite{James22} gave the detailed analysis for possible errors and concluded that works may produce erroneous results when assuming a 1-1 DM-z relation, which is consistent with our original intention to add errors in DM-z conversion.
Meanwhile, since our curve fitting is only for the situation that the error is not considered, the fitting result may have a slight change after the error is introduced, but this will not affect our final conclusion, because the statistical test values are given for the different errors.

\subsection{Magnetar origin of FRBs}
Next, we discuss the upper limit of the intrinsic luminosity that we obtained in Figure 1.
As concluded in our former work \citep{Cui21a} and combined with new CHIME data, the lognormal distribution is more supportive of the magnetar origin models \citep{Popov10, Lyutikov20}, and it is also valid in terms of the magnitude of the luminosity.
In particular, for 125 non-repeaters, the maximum luminosity ($L_{max}$) is about $2.86\times 10^{45}\,\rm  erg\, s^{-1}$ and the minimum luminosity ($L_{min}$) is about $6.83\times 10^{42}\,\rm  erg\, s^{-1}$ with the mean value of $2.42\times 10^{44}\,\rm  erg\, s^{-1}$.
Some giant flares from soft-$\gamma$-ray repeaters (SGRs) that are about $10^{44}\,\rm  erg\, s^{-1}$ in hard X-ray or soft $\gamma$-ray band \citep{Mazets79, Hurley99, Mereghetti05}.
If we consider a relatively high radio efficiency of 0.1 between the radio and X-ray luminosity, then the above SGR luminosity may conform to the FRBs' luminosity.
However, if the relative low radio emission efficiency is about $10^{-5}$, like SGR 1935+2154 \citep{Margalit20}, then the radio luminosity of SGR will not meet the above FRBs.
Only the rare huge-giant flare can satisfy with the obtained FRB luminosity \citep{Lyutikov17}, like the spike luminosity of about $10^{47}\,\rm  erg\, s^{-1}$ in the X-ray band \citep{Hurley05, Palmer05}, or the newborn millisecond magnetars may produce such huge-giant flares that have not been observed.

However, for the $L_{max}$ of FRBs being much higher than the inferred radio luminosity of the giant flares of SGRs as observed, it may be necessary to consider that the special magnetars have the super-strong magnetic fields \citep{Beloborodov17}, higher than the known values of $10^{15}\,\rm G$ \citep{Duncan92, Kaspi17}.
If their magnetic fields increase by half or one order of magnitude to $10^{16}\,\rm G$, the total released magnetic energy will increase by one or two orders of magnitude, as  $E_{tot}\sim B^2V/8\pi   \sim 10^{49}\,\rm erg$, where $B \sim 10^{16}\,\rm G$ is the surface magnetic field of a young magnetar and $V\sim 10^{18} cm^{3}$ is the volume of the magnetar.
If we assume the radio emission efficiency to be about $10^{-4}$ and lasting time of FRBs to be 10 milliseconds, then the corresponding FRB luminosity of non-repeater will rise to $10^{47}\,\rm  erg\, s^{-1}$. 
Therefore, this indicates that the cosmic FRBs are likely to come from the rare and violent bursts or interactions of magnetars.

\subsection{Difference of repeaters and non-repeaters}
The possible distinction between the repeaters and non-repeaters is briefly discussed in this subsection.
Although the luminosity distribution of repeaters is also lognormal \citep{Cui21a}, the two groups have different statistical distributions.
The average luminosity of repeaters ($3.85\times 10^{43}\,\rm  erg\, s^{-1}$) is about three times lower than that of non-repeaters ($1.31\times 10^{44}\,\rm  erg\, s^{-1}$).
This implies that they may come from a similar origin but different environments or emission mechanisms, such as the magnetars with the mediate magnetic field strengths or special structures.
For example, the magnetic field strength of repeater source is about $10^{14-15}\,\rm G$, while that of non-repeater may be higher than $10^{15}\,\rm G$.
In terms of the FRB origin models, there are two promising candidate forms: violent outburst from magnetar and supergiant pulse from strong magnetic neutron star \citep{Popov18, Katz20, Zhang20}.
Specifically, the huge-giant flares from ultra-strong magnetar ($\sim 10^{16}\,\rm G$) \citep{Lyubarsky14, Murase16, Beloborodov17} may be one of the explanations for a single burst of non-repeaters, and extremely high luminosity ($\sim 10^{47}\,\rm erg\,s^{-1}$) could lead to the long burst interval that it is considered as a non-repeater.
On the other aspect, the supergiant pulses from the extragalactic neutron stars \citep{Cordes16, Connor16} may be a good description for repeaters.
Consequently, our results further support the opinion of multiple origins of FRBs, which is also implied in previous studies based on CHIME/FRB catalog 1 in different parameters \citep{Pleunis21} and another database \citep{Cui21b}.

In the end, we need to clarify that the luminosity of two samples may have an overlap part, which means that some non-repeaters may be repeated in the future, like FRB 171019 \citep{Kumar19}.
In other words, the merger of the binary system \citep{Totani13, Kashiyama13, Mingarelli15} and catastrophic collision events \citep{Geng15, Dai16} may also be mixed in the non-repeater sample.
However, the proportion of these one-time collision events should be a small part, because the luminosity distribution of non-repeaters shows a single lognormal distribution.
Therefore, the non-repeater samples can also be divided into two groups: the true non-repeaters and repeaters as a single observed burst.
Finally, because of the statement in Section 4.1 and a fewer data of only 10 repeaters, the low p-values need to be treated carefully, and its statistical conclusion might be tentative.
Thus we look forward to having more such data to uncover the mysteries of FRBs.
~\\
~\\
{\bf A brief summary of conclusions:}
\begin{itemize}
\item We added the latest 10 localized FRBs to reconfirm the Macquart relation, indicating that this relation is still credible for the known 18 localized FRBs.
\item The luminosity of repeaters and non-repeaters are calculated based on the distance estimated by the Macquart relation, and their distribution conforms with the lognormal with the different mean values and derivations.
However, the lognormal features decrease as the transition error of DM-z increases until the error reaches 40\%, at where the lognormal features are weakened.
\item These implies that the two samples possibly come from similar origins, such as magnetar or strong magnetic NS.  
\end{itemize}

\section*{Acknowledgments}
This work is supported by the National Natural Science Foundation of China (Grant No. 11988101, No. U1938117, No. U1731238, No. 11703003, No. 11725313, and 12163001),
the International Partnership Program of Chinese Academy of Sciences grant No. 114A11KYSB20160008,
the National Key R\&D Program of China No. 2016YFA0400702, and the Guizhou Provincial Science and Technology Foundation (Grant No. [2020]1Y019).
The data underlying this article are available in the references below:
(1) The data of repeating and non-repeaters are taken from the database of CHIME/FRB Catalog 1, available at \url{https://www.chime-frb.ca/home}.
(2) The information of localized FRB is provided at the web \url{http://frbhosts.org/}

\newpage
\section*{Appendix A: Verification of the Macquart relation with updated 18 localized FRBs}
When the Macquart relation was given, only 8 localized FRBs were considered.
Now, we add the newly localized 10 FRBs, and a total of 18 data points are taken into account to verify the Macquart relation.
Since the M81 is too close to us \citep{Bhardwaj21}, FRB 20200120E is not listed in the above 18 data.
The fitted line of 18 data is given, with the goodness of fit as  0.75.
As shown in Figure 5, our fitted line (solid) is almost parallel to that of the Macqurat relation (dashed), and the deviation of two slopes is less than 6\%.
This indicates that the DM-z relation is still available, at least for the case that z is less than 0.7.
While the obvious difference between the two lines is reflected as a  fact that our fitting has an intercept value of  84.43 $\rm pc \, cm^{-3}$  with the vertical axis.
A possible reason for this gap is that the different DM data have been used.
Our $\rm DM_{excess}$ contain the $\rm DM_{host}$ and $\rm DM_{sur}$, but the $\rm DM_{cosmic}$ in the Macquart relation does not.
Meanwhile, the contribution of the Milk way galaxy halo does not consider in our analysis either.
Therefore, the gap value of 84.43 $\rm pc \, cm^{-3}$ may infer the DM in the host galaxy, surrounding medium, halo,  or both of them.
\begin{figure}
    \centering
    \includegraphics[width=6cm]{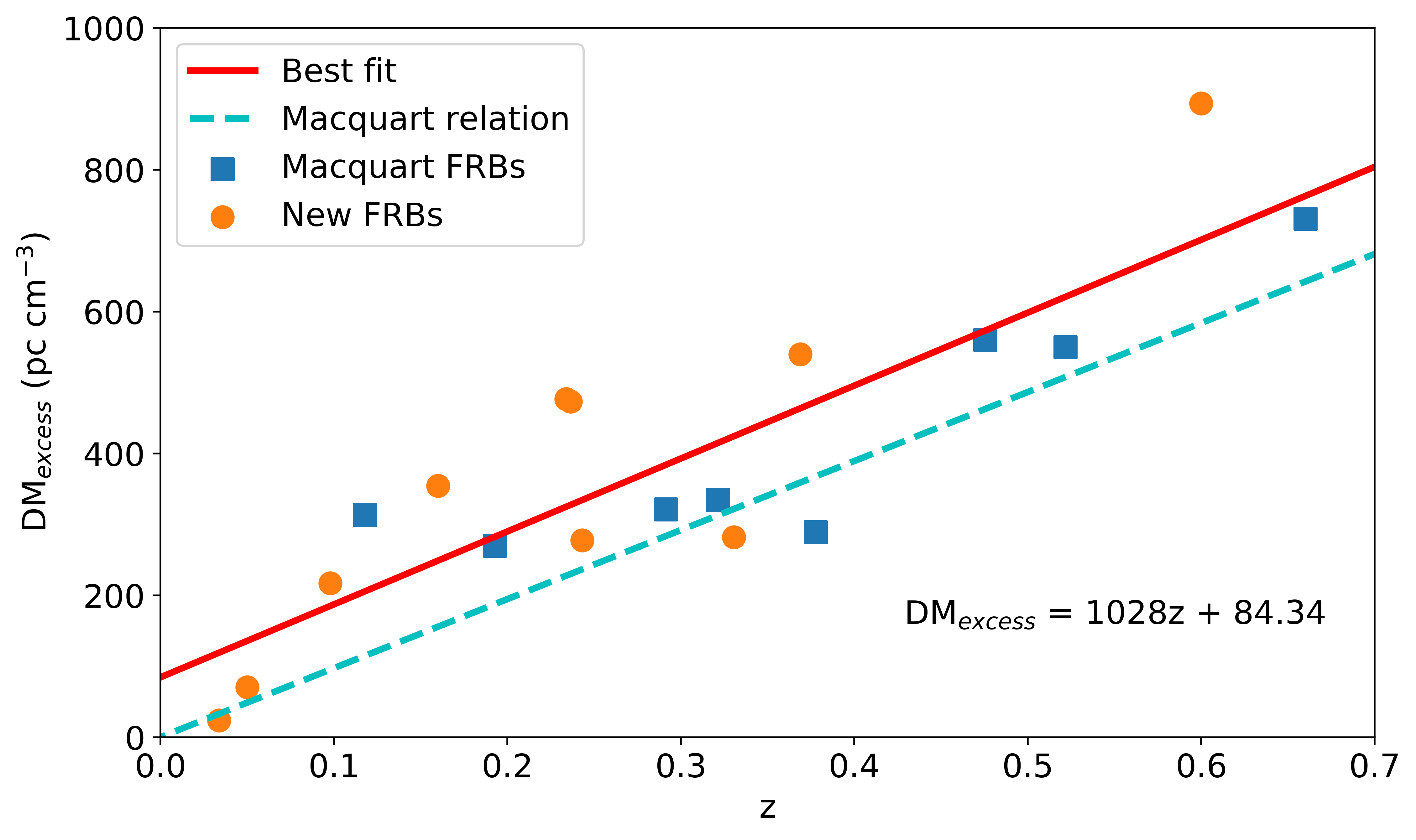}
    \caption{Diagram of dispersion measure (DM) and redshift (z).
    The solid line represents  the best fit curve of 18 localized FRBs, and the dashed line stands for  the Macquart relation.
    The squares represent 8 localized FRBs given by \citet{Macquart20}, and the 10 circles are the newly added data.}
    \label{fig5}
\end{figure}

\section*{Appendix B: Analysis of full DM data}
Under our data selection, we only analyze the repeaters and non-repeaters at low DM in the main text, so we discuss the full data here, including FRBs at both high and low DM.
Same as the above, we test lognormal feature for these date under K-S, Lilliefors, and A-D tests in Table 4.
The results are consistent with the low DM data. 
Meanwhile, the goodness of fit, K-S and M-W-W tests are also applied on the fitted curves, and the results are shown in Table 5 which further support the lognormal distribution of luminosity for non-repeaters.

Besides that, we plot full DM data in Figure 6 to compare whether repeaters and non-repeaters have the same distribution with mean values of the two samples (repeaters: $1.23\times10^{44}\,erg\,s^{-1}$, non-repeater: $5.48\times10^{44}\,erg\,s^{-1}$).
Although the repeaters and non-repeaters still show the different distributions ($p_{ks}=6.17\times10^{-4}$ and $p_{mww}=5.70\times10^{-5}$) like in the case of low DM selection, the mean values of all data are inconsistent with the data at low DM, indicating that the high and low DM may be different.
Furthermore, the maximum and minimum luminosity of non-repeaters is $8.00\times 10^{45}\,\rm  erg\, s^{-1}$ and $6.83\times 10^{42}\,\rm  erg\, s^{-1}$, respectively, which has a wider distribution range than the sample at low DM.
Therefore, these imply that the reasons for these differences are possibly due to the observational effects, data processing methods, or even their intrinsic physical properties, and we need further study to figure out the puzzles.

Overall, no matter whether we use the low DM or full DM data, it does not impact our main conclusions on the luminosity distribution form of non-repeaters, which is lognormal type is better for describing them.

\begin{table}
\centering
\caption{Statistical test results of lognormal distribution for repeaters and non-repeaters for full DM}
{
\begin{tabular}{@{}lcccc@{}}
\hline
\noalign{\smallskip}
\bf FRBs &\bf Number &\bf K-S test &\bf Lilliefors test &\bf A-D test$^a$ \\
\hline
\noalign{\smallskip}
Non-repeaters & 255 & 0.452 & 0.111 & (0.481, 0.775)\\
Repeaters & 18 & 0.716 & 0.238 & (0.397, 0.687)\\
\hline
\end{tabular}
}
\label{tab4}
\begin{flushleft}
Notes. $^a$ The results of A-D test are shown in the form of statistic and critical values (statistic, critical values).
\end{flushleft}
\end{table}

\begin{table}
    \centering
    \caption{Statistical test results of different luminosity distribution for all non-repeaters}{
    \begin{tabular}{@{}lccc@{}}
    \hline
    \noalign{\smallskip}
    \bf Different types&\bf Goodness of fit &\bf K-S test  &\bf M-W-W test \\
    \hline
    \noalign{\smallskip}
    Lognormal & 0.99994 & 0.851 & 0.670 \\
    Power-law & 0.614 & $4.57\times 10^{-4}$ & $3.29\times 10^{-3}$\\
    \hline
    \end{tabular}}
    \label{tab5}
    \begin{flushleft}
    \end{flushleft}
\end{table}
\begin{figure}
    \centering
    \includegraphics[width=6cm]{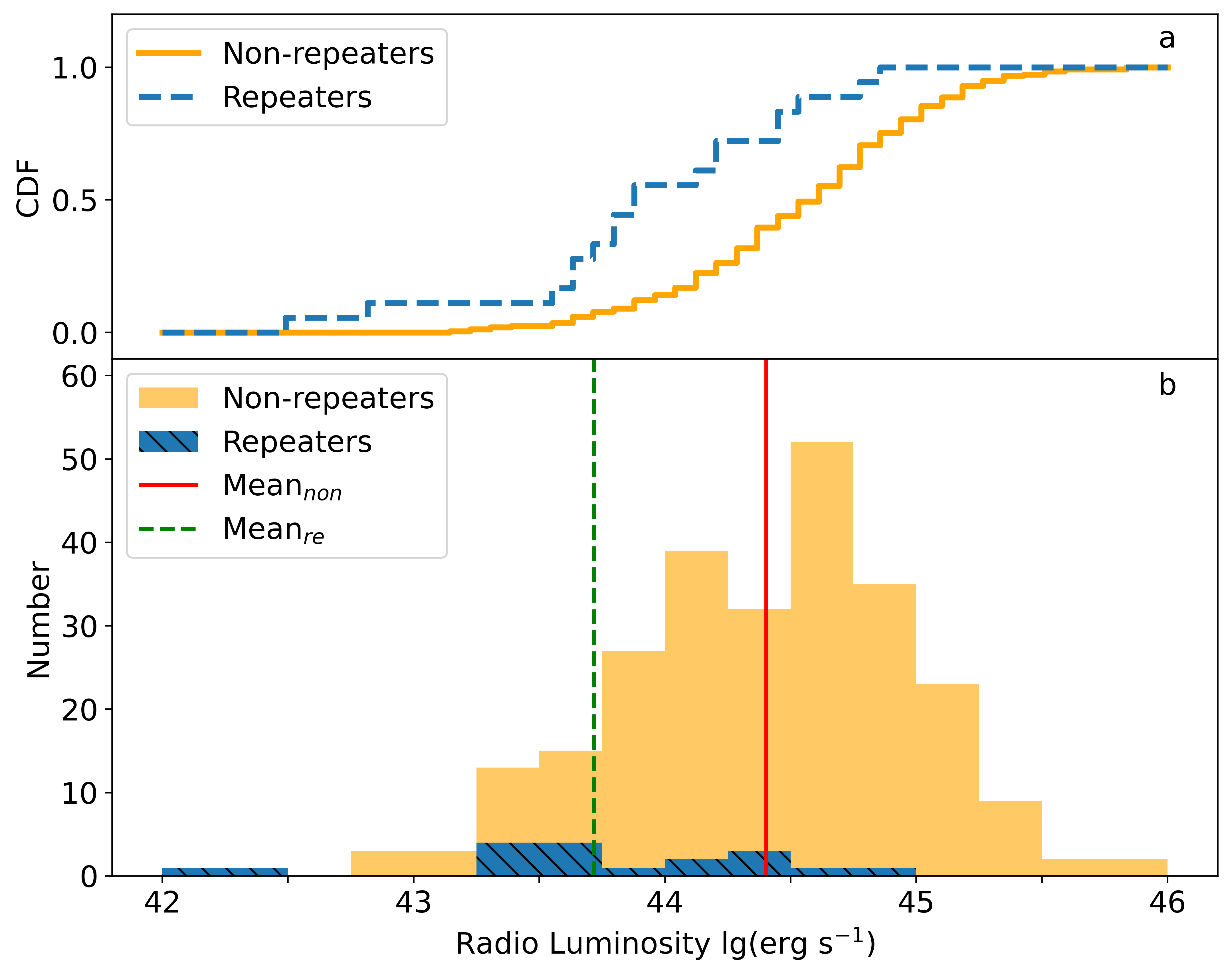}
    \caption{The distinction for all repeaters and non-repeaters (both low and high dispersion measure (DM)) in the aspect of luminosity.
    The top panel (sub-figure a) is the CDF of repeaters and non-repeaters for luminosity.
    The dashed (solid) line is for repeaters (non-repeaters).
    The bottom panel (sub-figure b) is the histogram of luminosity for two samples.
    The cross-hatched (empty) histogram means the repeaters (non-repeaters).
    The dashed (solid) line represents the mean value of repeaters (non-repeaters).}
    \label{fig6}
\end{figure}

\bibliography{CHIME-BIB}

\section*{Statements \& Declarations}
\subsection*{Funding}
This work was supported by the National Natural Science Foundation of China (Grant No. 11988101, No. U1938117, No. U1731238, No. 11703003 and No. 11725313),
the International Partnership Program of Chinese Academy of Sciences grant No. 114A11KYSB20160008,
the National Key R\&D Program of China No. 2016YFA0400702, and the Guizhou Provincial Science and Technology Foundation (Grant No. [2020]1Y019).

\subsection*{Data availability}
The dataset of repeaters and non-repeaters is available in the CHIME/FRB Catalog 1 repository, \url{https://www.chime-frb.ca/home}.
The dataset of localized FRBs is available in the FRB HOST DATABASE, \url{http://frbhosts.org/}.

\subsection*{Competing Interests}
The authors have no relevant financial or non-financial interests to disclose.

\subsection*{Author Contributions}
All authors contributed to the study conception and design. 
Xianghan Cui and Chengmin Zhang wrote the main manuscript text, Di Li and Jianwei Zhang provied analytical methods, Di Li, Bo Peng, Weiwei Zhu. and Richard Strom illustrated physical properties, Shuangqiang Wang, Na Wang and Qingdong Wu gave suggestions on figures, and Dehua Wang and Yangyi Yan modify the manuscript. 
All authors participated in the discussion and read the manuscript.

\label{lastpage}

\end{document}